\documentclass[fleqn,usenatbib]{mnras}

\usepackage{newtxtext,newtxmath}
\usepackage{lscape}

\usepackage[T1]{fontenc}
\usepackage{ae,aecompl}

\newbox\grsign \setbox\grsign=\hbox{$>$} \newdimen\grdimen \grdimen=\ht\grsign
\newbox\simlessbox \newbox\simgreatbox
\setbox\simgreatbox=\hbox{\raise.5ex\hbox{$>$}\llap
     {\lower.5ex\hbox{$\sim$}}}\ht1=\grdimen\dp1=0pt
\setbox\simlessbox=\hbox{\raise.5ex\hbox{$<$}\llap
     {\lower.5ex\hbox{$\sim$}}}\ht2=\grdimen\dp2=0pt
\def\simgreater{\mathrel{\copy\simgreatbox}}
\def\simless{\mathrel{\copy\simlessbox}}
\newbox\simppropto
\setbox\simppropto=\hbox{\raise.5ex\hbox{$\sim$}\llap
     {\lower.5ex\hbox{$\propto$}}}\ht2=\grdimen\dp2=0pt

\usepackage{color}

\usepackage{textcomp}
\def\pm{\textpm}

\usepackage{graphicx}	
\usepackage{amssymb}	

\title[Chemistry of GCs with APOGEE]{ The Chemical Compositions of
Accreted and {\it in situ} Galactic Globular Clusters According to
SDSS/APOGEE}
\author[D. Horta et al.]{
Danny Horta $^{1}$,\thanks{E-mail: 
D.HortaDarrington@2018.ljmu.ac.uk (KTS)}
Ricardo P. Schiavon $^{1}$,
J. Ted Mackereth $^{1,2}$,
Timothy C. Beers $^{3}$,
\newauthor
Jos\'e G. Fern\'andez-Trincado$^{4}$,
Peter M. Frinchaboy$^{5}$,
D. A. Garc\'ia-Hern\'andez $^{6,7}$,
\newauthor
Doug Geisler $^{8,9}$,
Sten Hasselquist $^{10,11}$,
Henrik J\"{o}nsson $^{12,13}$,
Richard R. Lane $^{14,15}$,
\newauthor
Steven R. Majewski $^{16}$,
Szabolcs M\'{e}sz\'{a}ros $^{17,18}$,
Christian Moni Bidin $^{19}$,
\newauthor
David M. Nataf $^{20}$,
Alexandre Roman-Lopes $^{21}$,
Christian Nitschelm, $^{22}$
\newauthor
J. Vargas-Gonz\'{a}lez$^{23}$,
Gail Zasowski$^{11}$
\\
$^{1}$Astrophysics Research Institute, Liverpool John Moores University, 146 Brownlow Hill, Liverpool L3 5RF, UK \\
$^{2}$School of Astronomy and Astrophysics, University of Birmingham, Edgbaston, Birmimgham B15 2TT, UK\\
$^{3}$Department of Physics and JINA Center for the Evolution of the Elements, University of Notre Dame, Notre Dame, IN 46556, USA\\
$^{4}$Instituto de Astronom\'ia y Ciencias Planetarias, Universidad de Atacama, Copayapu 485, Copiap\'o, Chile\\
$^{5}$ Department of Physics $\&$ Astronomy, Texas Christian University, Fort Worth, TX 76129, USA\\
$^{6}$Instituto de Astrof\'isica de Canarias (IAC), E-38205 La Laguna, Tenerife, Spain\\
$^{7}$Universidad de La Laguna (ULL), Departamento de Astrof\'isica, E-38206 La Laguna, Tenerife, Spain\\
$^{8}$Departamento de Astronom\'{i}a, Universidad de Concepci\'{o}n, Casilla 160-C, Concepci\'{o}n, Chile\\
$^{9}$Departamento de Astronom\'ia, Facultad de Ciencias, Universidad de La Serena. Av. Juan Cisternas 1200, La Serena, Chile\\
$^{10}$New Mexico State University, Las Cruces, NM 88003, USA\\
$^{11}$Department of Physics $\&$ Astronomy, University of Utah, Salt Lake City, UT 84112, USA\\
$^{12}$Materials Science and Applied Mathematics, Malm\"o University, SE-205 06 Malm\"o, Sweden\\
$^{13}$Lund Observatory, Department of Astronomy and Theoretical Physics, Lund University, Box 43, SE-22100 Lund, Sweden\\
$^{14}${Instituto de Astronom\'ia y Ciencias Planetarias, Universidad de
Atacama, Copayapu 485, Copiap\'o, Chile}\\
$^{15}$ Instituto de Astrof\'{i}sica, Pontificia Universidad Cat\'{o}lica de Chile, Av. Vicuna Mackenna 4860, 782-0436 Macul, Santiago, Chile\\
$^{16}$ Dept. of Astronomy, University of Virginia, Charlottesville, VA 22904-4325, USA\\
$^{17}$ELTE E\"otv\"os Lor\'and University, Gothard Astrophysical Observatory, 9700 Szombathely, Szent Imre H. st. 112, Hungary \\
$^{18}$MTA-ELTE Exoplanet Research Group, 9700 Szombathely, Szent Imre h. st. 112, Hungary\\
$^{19}$ Instituto de Astronom\'{i}a, Universidad Cat\'{o}lica del Norte, Av. Angamos 0610, Antofagasta, Chile\\
$^{20}$Department of Physics $\&$ Astronomy, The Johns Hopkins University, Baltimore, MD 21218, USA\\
$^{21}$Departamento de F\'{i}sica, Facultad de Ciencias, Universidad de La Serena, Cisternas 1200, La Serena, Chile\\
$^{22}$Centro de Astronom\'{i}a (CITEVA), Universidad de Antofagasta, Avenida Angamos 601, Antofagasta 1270300, Chile\\
$^{23}$Centre for Astrophysics Research, School of Physics, Astronomy and Mathematics, University of Hertfordshire, College Lane, Hatfield AL10 9AB, UK}
\date{Accepted XXX. Received YYY; in original form ZZZ}
\pubyear{2019}
\begin{document}
\label{firstpage}
\pagerange{\pageref{firstpage}----\pageref{lastpage}}
\maketitle
\begin{abstract}
Studies of the
kinematics and chemical compositions of Galactic globular clusters
(GCs) enable the reconstruction of the history of star formation,
chemical evolution, and mass assembly of the Galaxy. Using the
latest data release  (DR16) of the SDSS/APOGEE survey, we
identify 3,090 stars associated with 46 GCs. Using a previously
defined kinematic association, we break the sample down into
eight separate groups and examine how the kinematics-based
classification maps into chemical composition
space, considering only $\alpha$ (mostly Si and Mg) elements and Fe. Our results show that: (i) The loci of both \textit{in situ} and accreted subgroups in chemical space match those of their field counterparts; (ii) GCs from different individual accreted subgroups occupy the same locus in chemical space.  This could either mean that they share a similar origin or that they are associated with distinct satellites which underwent similar chemical enrichment histories; (iii) The chemical compositions of the GCs associated with the low orbital energy subgroup defined by Massari and collaborators is broadly consistent with an \textit{in situ} origin.  However, at the low metallicity end, the distinction between accreted and \textit{in situ} populations is blurred; (iv) Regarding the status of GCs whose origin is ambiguous, we conclude the following: the position in Si-Fe plane suggests an \textit{in situ} origin for Liller 1 and a likely accreted origin for NGC~5904 and NGC~6388.  The case of NGC~288 is unclear, as its orbital properties suggest an accretion origin, its chemical composition suggests it may have formed {\it in situ}.

\end{abstract}
\section{Introduction}

In a $\Lambda$CDM cosmology, galaxies grow in mass due to the process
of hierarchical assembly, whereby low-mass structures merge
together to form the galaxies we observe  in the local universe.
The signature of this process can be identified in the Milky Way,
in the form of halo stellar streams (\citealp[
e.g.,][]{Helmi1999,Belokurov2006,Ibata2016}), substructure in phase
space, such as the Gaia-Enceladus/Sausage system
(\citealp[GE/S,][]{Belokurov2018,Haywood2018,Helmi2018,Mackereth2018}), and
ongoing accretion, such as the Sagittarius dwarf spheroidal \citep[Sgr
dSph,][]{Ibata1994}.

The satellites involved in such merger events naturally brought
with them a cohort of globular clusters (GCs) \citep[
e.g.,][]{Penarrubia2009}, which survived the tidal interaction
with the central halo and today are an integral part of the Galactic
GC system. For decades, this has been the focus of various
studies \citep[e.g.,][]{Searle1978,Fall1985,Ashman1992,Brodie2006}, aiming at using
age, chemical composition, and phase-space information in order to,
on one hand, understand the origin of the Galactic GC system, and
on the other constrain the early history of mass assembly of the
Milky Way.  Key to that enterprise is to discern which of the
Galactic GCs were formed \textit{in situ} and which were accreted.
In the last decade, the availability of precise ages
(\citealp{Marin-Franch2009,Vandenberg2013}) has led to the discovery
of the bifurcation in the age-metallicity relation of Galactic GCs
(\citealp{Marin-Franch2009, Forbes2010,Leaman2013}), which, combined
with results from high-resolution hydrodynamical cosmological
simulations of Milky Way analogues, has further constrained the
origin the Galactic GC system \citep{Kruijssen2019a,Myeong2019}.
Furthermore, the advent of the Gaia survey \citep{Gaia2018}, and the
resulting precise 6D phase-space information have made possible a
much better characterization of the properties of the Galactic GC
system.

Along those lines, a recent study by \cite{Massari2019}
presented a new classification of the Galactic GC system in terms
of the kinematic properties of its members.  By studying their
distribution in integral of motions (hereafter, IOM) space, \cite{Massari2019}
established an association of each GC to one of the following main
groups: Main Disk (MD), Main Bulge (MB), Gaia Enceladus (GE),
Sagittarius (Sag), Helmi Streams (H99), Sequoia (Seq), Low Energy
(LE) and High Energy (HE).

The 16th data release of the Sloan Digital Sky Survey \citep[DR16][]{Ahumada2019} includes data for over 450k stars
from the APOGEE survey \citep{Majewski2017}, placing us in an
advantageous position to obtain detailed chemical-abundance information
for stars that are members of a significant fraction of the total
Galactic GC population
(\citealp{Meszaros2015,Schiavon2017,Meszaros2018,Masseron2019,Nataf2019}). Such data will substantially further our understanding of
the origin of the Galactic GC system, and in the process will help
constrain the assembly history of the Milky Way.  In this paper, we present
an examination of the chemical properties of the GC groups identified
by \cite{Massari2019}. Our goal is to check whether subgroups that are defined purely on the basis of orbital properties can also be distinguished in terms of their chemical properties.  In the process it is also possible to verify whether the chemical compositions are consistent with the star formation and chemical enrichment histories expected from the systems they are associated with.

This paper is structured as follows. In Section~\ref{data}, we briefly describe our data. In Section~\ref{method},
we describe the sample used and the criteria adopted to define GC
membership, and in Section~\ref{results}, we present the results obtained from the examination of the chemical properties of the
kinematically defined GC groups. Section~\ref{conclusion} summarizes
our results and conclusions.

\section{Data} \label{data} 

We use data from the sixteenth data release of SDSS-IV \citep{Ahumada2019}, which contains refined elemental abundances
(J\"{o}nsson et al.  2019, submitted.) from
the APOGEE-2 survey \citep{Majewski2017}, which is one of four
SDSS-IV \citep{Blanton2017} experiments.  APOGEE-2 is a near-infrared high-signal-to-noise ratio (S/N > 100 pixel$^{-1}$),
high-resolution (R $\sim$22,500) spectroscopic survey of over
450,000 Milky Way stars in the near-infrared \textit{H} Band (1.5--1.7
$\mu$m). Observations were based on two twin NIR spectrographs \citep{Wilson2019} attached to the 2.5~m telescopes at Apache Point \citep{Gunn2006}, and Las Campanas Observatories.  
Targets were selected in general from the 2MASS point-source catalogue,
employing a dereddened $(J - Ks)$ $\geq$ 0.3 colour cut in up to
three apparent $H$ magnitude bins. Reddening corrections were
determined using the Rayleigh-Jeans Colour
Excess method \citep[RJCE;][]{Majewski2011}, based on NIR photometry from the 2MASS point source catalogue \citep{Skrutskie2006}, and mid-IR data from the Spitzer-IRAC GLIMPSE-I,-II, and -3D \citep{Churchwell2009} when available from WISE \citep{Wright2010}.
A more in-depth description of the APOGEE survey, target selection, raw data, data reduction and spectral analysis pipelines can be found in
\citet{Majewski2017}, \citet{Holtzman2015}, \citet{Nidever2015},\citet{Perez2015},
\citet{Jonsson2018}, \citet{Zasowski2017},
respectively. All the APOGEE data products used in this paper are
those output by the standard data reduction and analysis pipeline.
The data are first processed (\citet{Nidever2015} $\&$ J\"{o}nsson
et al, in prep.) before being fed into the APOGEE Stellar Parameters
and chemical-abundances Pipeline \citep[ASPCAP;][ J\"{o}nsson
et al in prep]{Perez2015}. ASPCAP makes use of a specifically
computed spectral library (\citealp[][and J\"{o}nsson et al, in
prep]{Zamora2015,Holtzman2018}), calculated using a customised
\textit{H}-band line-list \citep[][Cunha et al., in prep.]{Shetrone2015}, from which the
outputs are analysed, calibrated, and tabulated \citep[][]{Holtzman2018}.

\section{Globular cluster sample and membership}
\label{method}
\subsection{Main sample}
\label{main_sampling}
In this subsection, we describe the method employed for determining
the GC sample in APOGEE DR16. We build on previous
work that has derived a GC sample in earlier data releases of APOGEE
( \citealp{Meszaros2015,Schiavon2017,Nataf2019})
and use the GC catalogues from \citet[][]{Harris1996},
\citet[][]{Baumgardt2018} and \citet{Baumgardt2019} in order to
establish GC membership of stars in the DR16 sample. The
methodology employed for identifying GC members is two-fold: the
first step comprises the determination of an initial sample based
on the values from the aforementioned catalogues. For this, we make
use of data on GC positions (Galactic latitude and longitude), radial velocities, the radial velocity dispersions,
tidal radii and mean metallicity values.

%

We use these values, and the values provided by APOGEE DR16
catalogue, to associate any star to be a member of a GC if:\\
\\
$i)$  $\bigl|$~[Fe/H]$_\star - \langle$[Fe/H]$_{GC}\rangle~\bigl|~\leq$~0.5\\ 
\\
$ii)$ $\bigl|$~\textit{rv}$_{\star} -
\langle$\textit{rv}$_{GC}\rangle~\bigl|~\leq$~$~2\,\sigma_{GC}$\\ 
\\
$iii)$ $d_{proj}~\leq~~2\,r_{vir}$\\


\noindent where [Fe/H] is the iron abundance, {\it $rv_\star$} is the
stellar heliocentric radial velocity, $\sigma_{GC}$ is the cluster's
radial velocity dispersion, $d_{proj}$ is the projected distance
between the star and the GC centre, and $r_{vir}$ is the cluster's
tidal radius. For GCs which are known to present a spead in metallicity (namely, NGC\,6715, Terzan\,5 and $\omega$~Cen), criterion \textit{i}) was ommitted.


The GC iron abundances and centre coordinates were extracted
from the 2010 edition of the Harris catalogue \citep{Harris1996},
whereas the GC radial velocities, velocity dispersions, and tidal radii were obtained from the latest versions of the
Baumgardt \& Hilker
catalogue\footnote{\url{http://physwww.mcmaster.ca/~harris/mwgc.dat}},\footnote{\url{https://people.smp.uq.edu.au/HolgerBaumgardt/globular/}}
(\citealt{Baumgardt2018,Baumgardt2019}).  The stellar
data come from APOGEE.\\

\begin{figure}
\center
  \includegraphics[width=7.5cm]{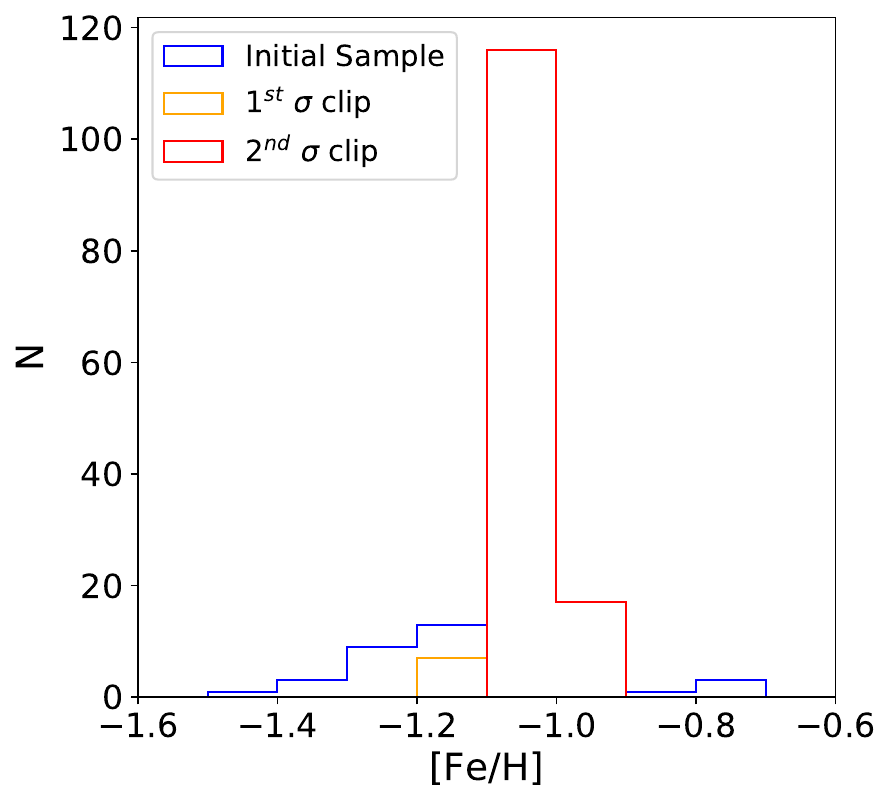}
  \caption{ Example of the $\sigma$-clipping method employed to determine
  GC candidate members from our initial sample. The mean metallicity
  value from the Harris catalogue for NGC\,6121 is [Fe/H] = --1.16,
  which lies very close to the peak of the distribution of our
  sample.}\label{fig:awesome_image3}
\end{figure}

The first step of the procedure consisted of the application of
criteria {\it (i)} to {\it (iii)} above, which yielded a preliminary
sample of $\sim$3,650 stars. Having obtained this preliminary sample,
the second step involved examining the metallicity distribution
functions (MDFs) of the candidates selected in the first pass, which
adopted a very broad [Fe/H] search interval. If the MDF
peaked at a value within 0.3 dex from the Harris catalogue value, and the distribution did not present tails of more than 0.3 dex away from the mean [Fe/H] value, all candidates
were deemed GC members.  For those cases in which the MDF peaked within 0.3 dex of the Harris catalogue value, but presented a broader,
less peaked, distribution, the sample was further cleaned through
$\sigma$-clipping, as follows. We computed the mean and
standard deviation of the [Fe/H] values, and then conducted the
$\sigma$-clipping procedure by removing any member candidates
that presented [Fe/H] abundances 1~$\sigma$ away from the mean.
However, for most GCs the resulting clipped MDF still presented tails in the distribution, so we had to perform a second $\sigma$ clipping, removing stars that deviated from the mean [Fe/H] value by more than 2 times the newly computed $\sigma$ value. Figure
~\ref{fig:awesome_image3} illustrates the $\sigma$-clipping method employed and how
successfully it works in defining GC candidate members from our
initial sample, removing any false positives. For a full visualization
of all the MDF cuts performed on the 43 GCs in our final main GC
sample and the resulting radial velocity distribution, see Appendix
~\ref{MDFS}.
We found that, by conducting this methodology, we were able to
minimise false positives and obtain a reliable sample of GC
members in APOGEE DR16. For 11 GCs from our original sample, fewer
than 3 star members could be identified, so these GCs were removed
from consideration.
The final sample contained 3,090 stars, associated with 46 GCs.
Our selection procedure is quite conservative and likely excludes
GC members.  However, sample purity is more important for our goals
than completeness. The mean elemental abundances, $rv$ values, and associated standard deviations for the member stars for each GC are listed in
Table \ref{tab1}. For GCs with large [Fe/H] spreads (namely, $\omega$~Cen, NGC\,6715 and Terzan\,5), mean abundances are not entirely meaningful. Moreover, in such cases the mean abundance ratios do not necessarily reflect those of the environments the GCs were born. Therefore, we removed these clusters from consideration. With this additional cut, our final working sample contains 1728 stars associated with 43 GCs.

Recently, \citet{Meszaros2019} performed a
careful GC membership analysis, obtaining a sample that is very
similar to ours.  They proceeded to study star-to-star internal
abundance variations in GCs, with data based on the BACCHUS
\citep{Masseron2016} abundance pipeline.  We have repeated our
entire analysis adopting both \citep{Meszaros2019} member
sample and abundances and obtained the same results as presented
in this paper.

\setlength{\tabcolsep}{20pt}
\begin{table*}
\label{tab:one}
\centering
\begin{tabular}{ |p{0.15cm}|p{1.3cm}|p{1.25cm}|p{1.3cm}|p{0.15cm}|p{1.3cm}|p{1.15cm}|p{1.4cm}}
\hline
 Name &  <[Fe/H]> & <[Si/Fe]>&<V$_{los}$> & Name & <[Fe/H]> & <[Si/Fe]> & <V$_{los}$> \\
\hline
\hline
NGC\,104 & --0.72\pm0.04 & 0.23\pm0.04& --18.8\pm7.3  & NGC\,6397 & --2.02\pm0.04  &0.3\pm0.06 & 19.5\pm2.8\\
\hline
NGC\,288 & --1.26\pm0.04 & 0.29\pm0.03& --44.5\pm2.2  &NGC\,6441 & --0.39\pm0.12  &0.13\pm0.13& 10.8\pm15.3 \\
\hline
NGC\,362 & --1.09\pm0.04  & 0.13\pm0.04& 223.6\pm5.3&NGC\,6522 & --1.04\pm0.06 &0.22\pm0.10& --12.8\pm7.6  \\
\hline
NGC\,1851 & --1.07\pm0.04  &0.14\pm0.05 & 320.7\pm5.9&NGC\,6539 & --0.39\pm0.09 &0.2\pm0.07& 33.8\pm4.4 \\
\hline
NGC\,1904 & --1.5\pm0.07  & 0.15\pm0.04& 207.4\pm2.6&NGC\,6540 & --1.01\pm0.03 &0.21\pm0.04 & --14.4\pm1.1 \\
\hline
NGC\,2808 & --1.04\pm0.06& 0.15\pm0.06 & 103.4\pm8.4 &NGC\,6544 & --1.44\pm0.05 &0.22\pm0.04& --38.6\pm4.6 \\
\hline
NGC\,3201 & --1.35\pm0.05& 0.16\pm0.04 & 495.4\pm3.3 &NGC\,6553 & --0.16\pm0.1 &0.09\pm0.07& 0.2\pm9.5 \\
\hline
NGC\,4590 & --2.24\pm0.07  & 0.33\pm0.05& --94.0\pm3.2&NGC\,6656 & --1.69\pm0.05 &0.29\pm0.11& --146.6\pm5.6  \\
\hline
NGC\,5024 & --1.92\pm0.04 & 0.25\pm0.08& --60.8\pm3.4 &NGC\,6715 & --- &---& ---\\
\hline
NGC\,5053 & --2.15\pm0.15 & 0.38\pm0.09& 42.9\pm1.3 &NGC\,6723 & --1.0\pm0.06 &0.25\pm0.03& --93.5\pm3.4 \\
\hline
$\omega$~Cen & --- & ---& ---&NGC\,6752 & --1.47\pm0.03 & 0.25\pm0.05 & --26.3\pm4.7\\
\hline
NGC\,5272 & --1.4\pm0.06  &0.17\pm0.07& --146.2\pm4.1 &NGC\,6760 & --0.71\pm0.1  & 0.19\pm0.05& --1.5\pm5.8\\
\hline
NGC\,5466 & --1.78\pm0.06 & 0.19\pm0.12& 108.1\pm1.0 &NGC\,6809 & --1.75\pm0.04  &0.23\pm0.05& 176.1\pm3.6 \\
\hline
NGC\,5904 & --1.19\pm0.05 &0.18\pm0.05 & 53.8\pm4.9&NGC\,6838 & --0.73\pm0.04 &0.22\pm0.03& --22.7\pm2.1 \\
\hline
NGC\,6121 & --1.04\pm0.03 & 0.34\pm0.04& 70.9\pm3.4 &NGC\,7078 & --2.28\pm0.05  &0.31\pm0.07 & --104.3\pm5.2\\
\hline
NGC\,6171 & --0.97\pm0.06  & 0.32\pm0.08& --33.8\pm3.1&NGC\,7089 & --1.46\pm0.06  &0.19\pm0.07& --3.6\pm5.5 \\
\hline
NGC\,6205 & --1.46\pm0.04 & 0.21\pm0.07& --246.3\pm5.2 &Terzan\,2 & --0.82\pm0.05  &0.26\pm0.02 & 133.2\pm1.4\\
\hline
NGC\,6218 & --1.26\pm0.03  & 0.26\pm0.05& --40.7\pm3.1&Terzan\,5 & ---  &--- & ---\\
\hline
NGC\,6229 & --1.25\pm0.05 & 0.19\pm0.06& --137.8\pm2.7&Pal\,5 & --1.24\pm0.02  &0.11\pm0.04& --58.9\pm0.5 \\
\hline
NGC\,6254 & --1.49\pm0.04 & 0.27\pm0.05& 75.8\pm3.9 &Pal\,6 & --0.81\pm0.09  & 0.28\pm0.07& 172.9\pm2.3\\
\hline
NGC\,6341 & --2.22\pm0.03  & 0.31\pm0.07& --118.2\pm6.7&Pal\,10 & 0.09\pm0.06  &0.0\pm0.01& --32.7\pm4.9 \\
\hline
NGC\,6380 & --0.72\pm0.05  & 0.21\pm0.02& --1.8\pm7.8&Liller\,1 & --0.03\pm0.05  & 0.01\pm0.05& 61.8\pm3.5\\
\hline
NGC\,6388 & --0.54\pm0.06  & --0.03\pm0.1& 80.1\pm10.5&HP\,1 & --1.14\pm0.07  & 0.22\pm0.06& 40.9\pm4.8\\
\hline
\hline
\end{tabular}
\caption{From left to right, GC name, mean [Fe/H], mean [Si/Fe], and mean radial velocity obtained for the final list of GCs in the main GC sample from APOGEE DR16.}
\label{tab1}
\end{table*}
\setlength{\tabcolsep}{20pt}
\begin{table*}
\label{tab:two}
\centering
\begin{tabular}{ |p{.6cm}|p{1.6cm}|p{1.6cm}|p{.6cm}|p{1.6cm}|p{1.6cm}}
\hline
 Name & E  &L$_{Z}$   &  Name &   E &  L$_{Z}$  \\
 &[km$^{2}$/s$^{2}$] &[10$^{3}$ kpc * km/s]  &&[km$^{2}$/s$^{2}$] &[10$^{3}$ kpc * km/s]  \\
\hline
\hline
NGC\,104 & --50892$^{+74}_{-94}$&0.62$^{+0.01}_{-0.01}$ & NGC\,6397 &--66705$^{+193}_{-190}$ & 0.33$^{+0.01}_{-0.01}$  \\
\hline
NGC\,288 & --41116$^{+1125}_{-947}$ & --0.34$^{+0.03}_{-0.03}$& NGC\,6441& --93869$^{+2812}_{-3479}$ & 0.21$^{+0.03}_{-0.03}$  \\
\hline
NGC\,362 &--41406$^{+1436}_{-1391}$& --0.07$^{+0.01}_{-0.01}$&NGC\,6522 &--129873$^{+1043}_{-737}$ & 0.01$^{+0.01}_{-0.01}$ \\
\hline
NGC\,1851 &--21934$^{+537}_{-521}$&--0.22$^{+0.02}_{-0.04}$&NGC\,6539 & --88145$^{+1367}_{-1378}$ &  --0.18$^{+0.00}_{-0.00}$\\
\hline
NGC\,1904 &--26641$^{+655}_{-650}$&--0.17$^{+0.05}_{-0.04}$ &NGC\,6540 & --108006$^{+1814}_{-2001}$ & 0.18$^{+0.01}_{-0.01}$\\
\hline
NGC\,2808 &--39859$^{+541}_{-690}$&0.13$^{+0.01}_{-0.01}$&NGC\,6544 &--85401$^{+474}_{-386}$& --0.07$^{+0.02}_{-0.02}$ \\
\hline
NGC\,3201 &6964$^{+1307}_{-1402}$&--1.65$^{+0.02}_{-0.02}$&NGC\,6553 & --101735$^{+2577}_{-1883}$ & 0.25$^{+0.02}_{-0.02}$ \\
\hline
NGC\,4590 &--1010$^{+1311}_{-801}$&1.27$^{+0.02}_{-0.01}$&NGC\,6656 &--49027$^{+204}_{-294}$ & 0.47$^{+0.01}_{-0.01}$  \\
\hline
NGC\,5024 &--14407$^{+851}_{-861}$ &0.33$^{+0.01}_{-0.02}$&NGC\,6715 & 22391$^{+3784}_{-3525}$ & 0.79$^{+0.01}_{-0.01}$ \\
\hline
NGC\,5053 &--18934$^{+925}_{-645}$&0.29$^{+0.02}_{-0.02}$ &NGC\,6723 & --85738$^{+958}_{-1010}$ & --0.02$^{+0.01}_{-0.01}$ \\
\hline
$\omega$~Cen & --66584$^{+413}_{-271}$ &--0.35$^{+0.01}_{-0.01}$ &NGC\,6752 &--69155$^{+484}_{-522}$ & 0.41$^{+0.01}_{-0.01}$ \\
\hline
NGC\,5272 &--26985$^{+454}_{-549}$&0.42$^{+0.01}_{-0.01}$  & NGC\,6760 & --78723$^{+322}_{-265}$&0.34$^{+0.01}_{-0.01}$ \\
\hline
NGC\,5466 &18295$^{+2946}_{-3170}$&--0.54$^{+0.04}_{-0.04}$ &NGC\,6809 & --69856$^{+220}_{-212}$&0.09$^{+0.01}_{-0.01}$ \\
\hline
NGC\,5904 &--4945$^{+2821}_{-2685}$&0.16$^{+0.01}_{-0.01}$& NGC\,6838 &--56897$^{+99}_{-92}$& 0.67$^{+0.00}_{-0.00}$\\
\hline
NGC\,6121 &--78561$^{+289}_{-377}$&--0.07$^{+0.02}_{-0.02}$ &NGC\,7078 & --44942$^{+811}_{-912}$ & 0.55$^{+0.02}_{-0.02}$ \\
\hline
NGC\,6171 &--93415$^{+343}_{-259}$ & 0.06$^{+0.01}_{-0.01}$  &NGC\,7089 & --24793$^{+1316}_{-1645}$ & --0.15$^{+0.04}_{-0.03}$ \\
\hline
NGC\,6205 &--55754$^{+314}_{-363}$&--0.2$^{+0.01}_{-0.01}$&Terzan\,2 & --142240$^{+1564}_{-1992}$ & --0.04$^{+0.00}_{-0.00}$  \\
\hline
NGC\,6218 &--9536$^{+878}_{-819}$&0.06$^{+0.11}_{-0.13}$ &Terzan\,5 &  --139676$^{+2229}_{-2447}$ & --0.02$^{+0.01}_{-0.01}$ \\
\hline
NGC\,6229 &--77618$^{+518}_{-442}$&0.24$^{+0.01}_{-0.01}$ &Pal\,5 &--9666$^{+3131}_{-2722}$&0.88$^{+0.08}_{-0.08}$ \\
\hline
NGC\,6254 &--76969$^{+692}_{-537}$&0.25$^{+0.01}_{-0.01}$& Pal\,6 & --96285$^{+1413}_{-1744}$&--0.01$^{+0.00}_{-0.01}$\\
\hline
NGC\,6341 &--49372$^{+809}_{-728}$&--0.05$^{+0.01}_{-0.01}$ &Pal\,10 &--59725$^{+733}_{-648}$&0.6$^{0.01}_{-0.01}$ \\
\hline
NGC\,6380 & --105529$^{+3090}_{-2754}$ & --0.02$^{+0.01}_{-0.01}$&Liller\,1 & -- & -- \\
\hline
NGC\,6388 & --101561$^{+1853}_{-1486}$&--0.13$^{+0.01}_{-0.01}$ &HP\,1 & --114308$^{+5219}_{-3012}$&--0.01$^{+0.01}_{-0.01}$\\
\hline
\hline
\end{tabular}
\caption{From left to right, GC name, mean orbital energy, and mean angular momentum obtained for the final list of GCs in the main GC sample from APOGEE DR16 using the \texttt{MWPotential2014} \citep{Galpy2015}. There is no 6D phase-space information for Liller\,1 provided in \citealp[][]{Vasiliev2019}, thus we are unable to obtain IOM for this GC.}
\label{tab2}
\end{table*}

\subsection{Globular cluster groups}

\begin{figure}
  \includegraphics[width=\linewidth]{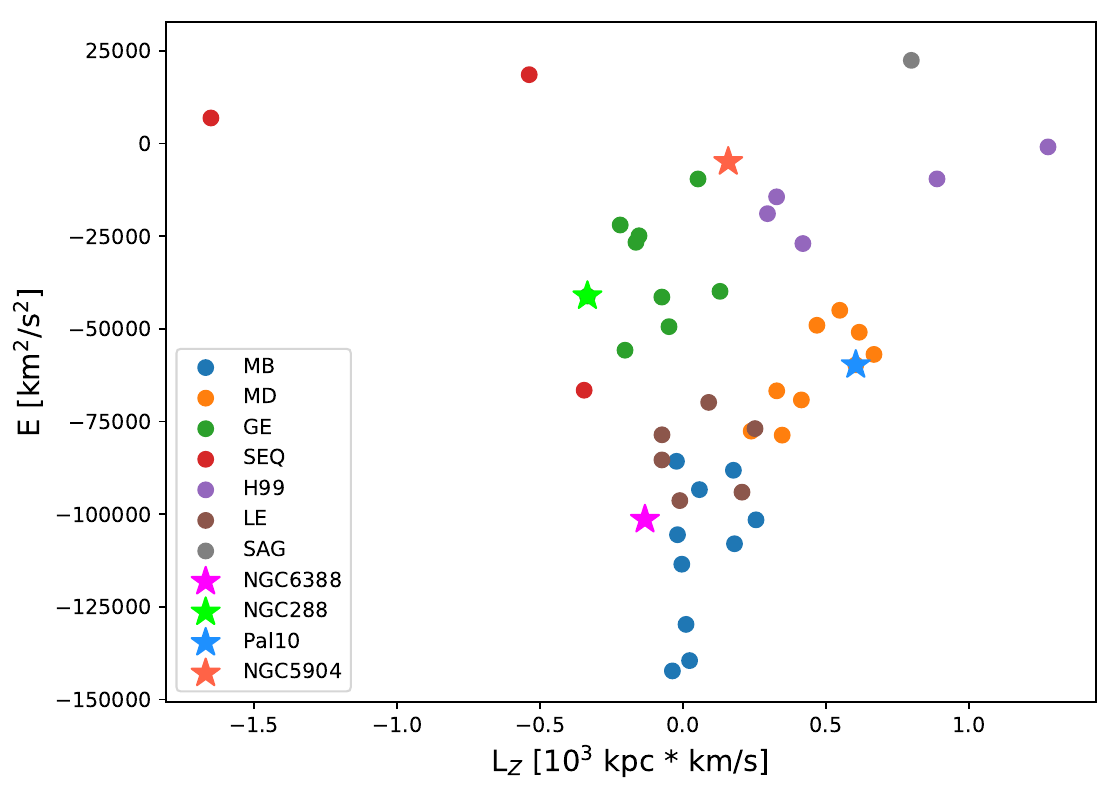}
  \caption{Orbital energy and vertical action as a function of orbital azimuthal action for the 46 GCs obtained in our initial main sample, divided into the kinematic associations identified by \citet[][]{Massari2019}.}
  \label{kinematics}
\end{figure}

In this subsection we briefly discuss how the GCs in our sample are
distributed across the kinematic groups defined by \cite{Massari2019}
(see Fig ~\ref{kinematics}).  From our main sample of 43 GCs,
we find that 9 can be associated to the MD group, 10 to the MB, 9
to the GE dwarf spheroidal, 5 to the H99, 6 to the LE group and 1 to
the Seq dwarf spheroidal.  An
additional 5 GCs from our sample could not be unambigously associated
to a single group {by \cite{Massari2019}}. The GC NGC~3201 could be associated to either of the
GE or Seq group, NGC~5904 could belong to either GE or H99, and
Liller~1 is listed as unclassified. Similarly,
NGC\,6388 is originally classified as a MB GC by \citep{Massari2019}.
However, recent work has shown that NGC\,6388 can be associated to
the Sequoia accretion event based on its eccentric-retrograde
orbit \citep[][]{Myeong2019}. For this work, we initially follow the
\citet{Massari2019} classification and include NGC\,6388 in the MB
subgroup, and study its chemical-abundances in order to discern if
this GC is from \textit{in situ} or accreted origin. Along the same
lines, since it has recently been shown that
NGC~3201 could be associated kinematically to the Sequoia dwarf
remnant \citep{Myeong2019}, we choose to include these GCs in the
Seq group.  The remaining two GCs for which \cite{Massari2019} do
not find a clear subgroup association (namely, NGC~5904 and Liller~1)
are initially marked as unclassified and are discussed in
Section~\ref{unclassified}. For the final list of the GCs obtained
in the main sample and the kinematic group association see Table
\ref{tab3}.

\setlength{\tabcolsep}{18pt}

\begin{table*}
\label{tab:associations}
\begin{tabular}{ |p{7.5cm}|p{7.5cm}|  }
\hline
 Kinematic Group          &  Associated GCs \\
\hline
\hline
  Main-Disc &  NGC\,7078(30), NGC\,6760(11), NGC\,6838(37), NGC\,6218(62), NGC\,6397(46), NGC\,6752(97), NGC\,104(176), NGC\,6656(35), Pal\,10(3)\\
   \hline
  Main-Bulge & NGC\,6539(6), NGC\,6171(51), Terzan\,2(4), NGC\,6553(23),  NGC\,6380(15), NGC\,6522(6),  \textbf{NGC\,6388(24)}, NGC\,6540(4), NGC\,6723(7), HP\,1(12)\\
   \hline
   Gaia-Enceladus & NGC\,1904(17), NGC\,2808(66), NGC\,6205(80), NGC\,6229(6), NGC\,6341(10), NGC\,362(49), NGC\,7089(26), NGC\,1851(30), NGC\,288(35), \textbf{NGC\,5904(167)} \\
   \hline
    Sequoia & NGC\,5466(7), NGC\,3201(114),  \textbf{NGC\,6388(24)} \\
   \hline
    Sagittarius &  ---\\
   \hline
    Helmi streams & NGC\,5024(18), NGC\,5053(11), NGC\,4590(13), NGC\,5272(110), Pal\,5(3), \textbf{NGC\,5904(167)} \\
   \hline
   Low-Energy & NGC\,6809(60), Pal\,6(5), NGC\,6441(28), NGC\,6121(140), NGC\,6254(59), NGC\,6544(21) \\
   \hline
    High-Energy & -- \\
   \hline
   XXX & \textbf{Liller\,1(4)}\\
  \hline
\hline
\end{tabular}
\caption{GCs obtained in APOGEE DR16 associated to the kinematic
subgroups as defined in \citet{Massari2019}, after removing GCs
with less than 3 star members. The GCs highlighted in bold are
associations that are uncertain. The number of APOGEE member stars
associated to each GC are given in parentheses.}
\label{tab3}
\end{table*}

\subsection{Elemental abundances and orbital parameters} 
In this paper, we report an examination of the APOGEE DR16
chemical-compositions for GCs from various subgroups. Specifically,
we focus on studying trends in  $\alpha$-element abundances as a
function of [Fe/H] to gain insights into the nature of the subgroups.
Our goal is to examine how the kinematic classification by
\citet{Massari2019} maps into chemical composition space.  In so
doing we expect to constrain the nature of the progenitors
of the various sub-systems making up the Galactic halo, given
the relation between chemical compositions stellar populations and
their histories of star formation and chemical enrichment.  This also makes possible a more clear distinction between
GCs formed {\it in situ} from those belonging to accreted systems

We focus on calibrated abundances \citep{Jonsson2018}, which have been compared in detail with independent determinations by other groups.  Of relevance to this work, \citet{Jonsson2018} show that Si abundances, although differing from those of some of the other groups by statistically significant zero-point shifts, show no trends with stellar parameters.  Since our results depend fundamentally on relative differences between abundances from a homogeneous set, such small zero-point effects are not important. The $\alpha$-element of choice for this study is silicon. We use
the [Si/Fe] abundances, as silicon has
been shown in previous data releases to be one of the most reliable
$\alpha$-abundance measurements in APOGEE \citep{Jonsson2018}.
Magnesium is another $\alpha$-element for which APOGEE provides
exquisite abundances, however it is affected by internal GC evolution
\citep[e.g.,][]{Bastian2018}, so we remove it from consideration
when using the main sample. In order to verify that our choice of
$\alpha$-element does not affect our conclusions, we performed the
analysis adopting [Mg/Fe] from first population stars and found
that our results are unchanged. In addition, we compared our mean [Si/Fe]
with those from the compilation by \cite{Pritzl2005}, finding our
values to be slightly lower, of the order of $\sim$ 0.1 dex. Again, such
a small zero-point difference has no impact on our results. 

Orbital parameters were estimated for our sample of
GCs as follows. We calculated the action integrals for each
of the 46 GCs using the potential defined by \citet[][\texttt{MWPotential2014}]{Galpy2015}, using the publicly
available code \texttt{galpy}
\footnote{\url{http://github.com/jobovy/galpy}}
(\citealp{Galpy2015,Galpy2018}). In order to obtain reliable kinematic
measurements, we draw 100 samples for each of the 6-D phase-space
parameters given by the GC table in \citet{Vasiliev2019}, and obtain
100 estimates of the orbital parameters for each cluster, for which
we then take the median and standard deviation as our value and
associated uncertainty.  Fig.~\ref{kinematics} displays the energy
(E) values obtained using this method as a function of the azimuthal
action ($L_{Z}$) for all the GCs in our main sample, colour-coded by subgroup association. Highlighted as star symbols are GCs
which display peculiar [Si/Fe] when compared to the remaining GCs
in the same subgroup. We find that our orbital energy values
differ from those of \citet{Massari2019}.  Such differences can be
traced back to the adoption of different Galactic potentials with
different total masses---while \cite{Massari2019} adopted a McMillan
potential \citep{McMillan2017}, the one adopted in this work was \texttt{MWPotential2014} \citep{Galpy2015}.  We assessed the impact of
Galactic potential choice on our results by re-running the calculations
using the McMillan potential, and found that the GC associations to the
various subgroups were unchanged, and are consistent with those found in \citet{Massari2019}.


\section{Results} \label{results}

\subsection{Disc, Bulge and Low Energy GCs}
\label{DiscBulge}

\begin{figure}
\label{tabletwo}
    \centering
    \includegraphics[width=1.\columnwidth]{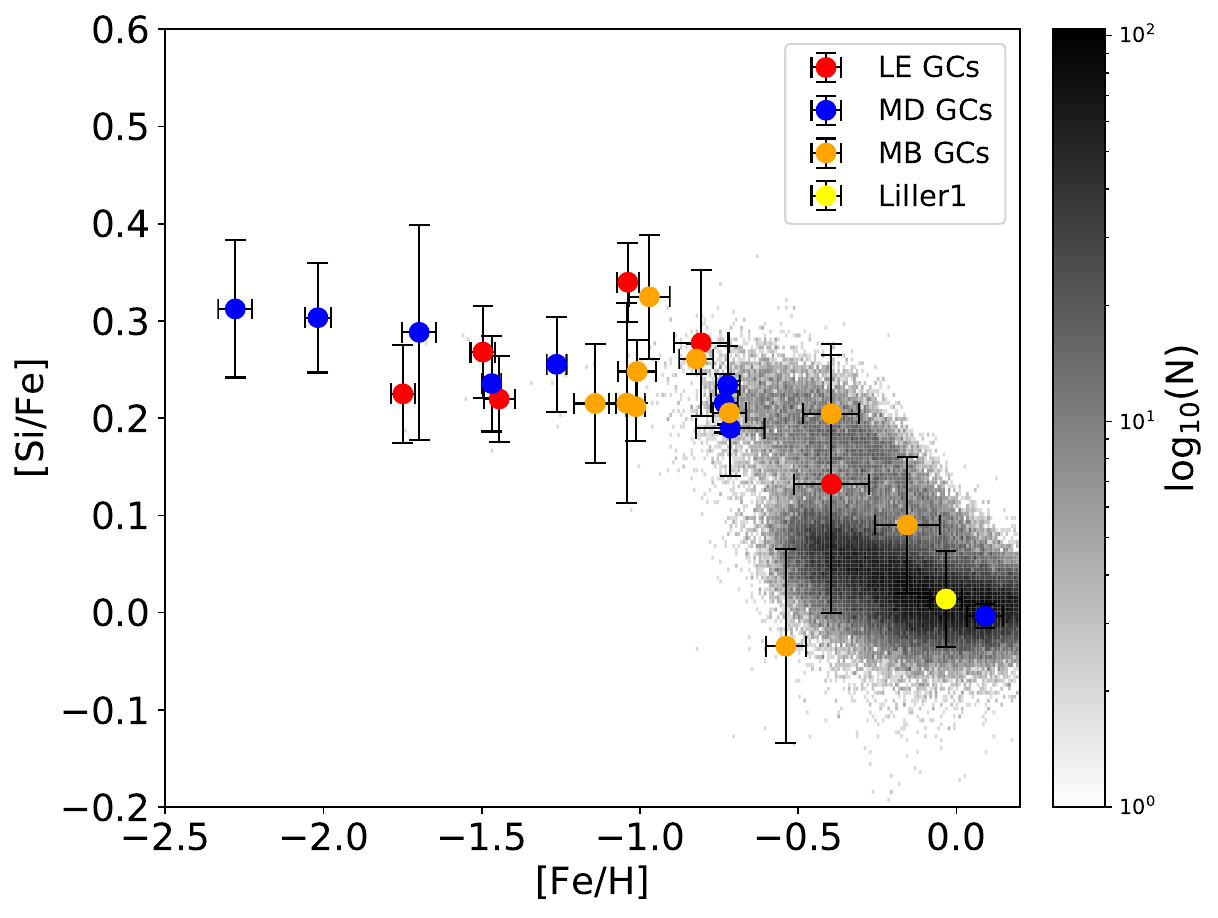}
    \caption{Mean [Si/Fe] vs [Fe/H] chemical-abundances for the Low Energy (red), Main Bulge (orange) and Main Disc (blue) GC subgroups, illustrated alongside Liller\,1 (yellow), with the
		    1~$\sigma$ spread represented in black error bars. In grey we show the Galactic disc and bulge field populations defined kinematically according to \citet{Massari2019}. From
		    these abundance plots, by accounting for the 1~$\sigma$ spread uncertainties, we find that the more [Fe/H] rich LE GCs, namely NGC\,6121, NGC\,6441 and Pal\,6 can be categorized to be from \textit{in situ} origin. The other three LE GCs still occupy the same locus as the MD/MB subgroups, however due to their low [Fe/H] abundances and position in the IOM space (Fig ~\ref{kinematics}), it is possible that these more metal-poor GCs could be from an accreted origin. Furthermore, we find that Liller\,1 occupies the same locus as the \textit{in situ} GCs, which coupled with its high [Fe/H] value can be classified as a MB GC.}
\label{SiFe_insitu}
\end{figure}

In Fig.~\ref{SiFe_insitu} we show the mean [Si/Fe] chemical-abundance
measurements as a function of [Fe/H] for the MD (blue symbols),
MB (orange symbols) and LE (red symbols) kinematically identified
subgroups. Also plotted are the data for the GC Liller~1
(yellow dot), which is discussed in Section~\ref{unclassified}.
At first glance the three subgroups occupy roughly the same
locus in [Si/Fe] space, resembling the region of abundance space
occupied by field stars from the disc and bulge components of the
Milky Way \citep[e.g.,][]{Hayden2015}.  Within the errors, the MD
population displays a low-metallicity [Si/Fe] plateau until reaching [Fe/H]
abundance values of approximately [Fe/H]~$\simeq-$0.6, for which
according to the Milky Way's Disc field population, we would expect
a knee towards lower [Si/Fe] values \citep{Alves2010}. When
considered in aggregate, the three subgroups display a clear knee
at about [Fe/H]$\sim$--0.8, with a plateau at [Si/Fe]$\sim$+0.25
at lower metallicities and a trend of decreasing [Si/Fe] for
increasing [Fe/H] at [Fe/H]$\simgreater$--0.8, which mimics the
behaviour of field stars.  One GC deviates clearly from this trend, namely NGC\,6388, with very low [Si/Fe]$\sim$0.0 at
[Fe/H]$\sim$-- 0.5.  We discuss this interesting GC separately in
Section~\ref{ngc6388}.

When the three subgroups are considered separately, however,
the relatively small number of GCs in our sample prevents the
unequivocal identification of a ``knee'' in the Si-Fe plane for any
of the subgroups in Figure~\ref{SiFe_insitu}.  In the case of the
MB subgroup, the sample does not contain enough GCs at [Fe/H] $\simless$ -- 0.8
to firmly establish the existence of a low metallicity [Si/Fe]
plateau.  The LE subgroup straddles properly both [Fe/H] sides of
the ``knee'' and the GCs seem to follow the same trend as the field
population, but the sample is too small for a solid conclusion.
The sample for the MD subgroup covers a wide range in [Fe/H] towards the metal-poor side of
the knee, but contains only one GC on the metal-rich end, whose
position on the Si-Fe is consistent with the existence of a knee
in that subgroup.  Again, the sample is not large enough at
[Fe/H]$\simgreater$-- 0.8 for a robust conclusion.  The GC on the
metal-rich end of the MD subgroup is Pal 10.  We checked to see
whether the orbital properties of this cluster match those of the
MD GC population. In Fig~\ref{kinematics} we show that Pal\,10 does
follow a disc-like orbit, displaying energy values of E$\sim$--6000
km$^{2}$/s$^{2}$ and following a prograde orbit (i.e. L$_{Z}\sim$0.7
10$^{3}$ kpc km/s), therefore it is likely to belong to the MD
subgroup.  All in all, the MB and MD subgroups follow the trend defined by
the field population, thus we conclude that these subgroups share an
{\it in situ} origin.


Since the origin of the LE subgroup is contentious, the locus
of GCs from that subgroup merits some attention. Our results show
that the GCs from this subgroup occupy the same locus in
[Si/Fe] vs [Fe/H] space as the MD/MB GCs. This result is in
line with the similarity of these subgroups in E-L$_{Z}$ space,
\citep[see Fig.~\ref{kinematics} of this paper and Fig.~3 of][]{Massari2019}
We note, however, that at [Fe/H]~$\simless-1.5$ it is almost
impossible to distinguish between accreted and {\it in situ} GCs
in the Si-Fe plane, so that an accreted origin for the three most
metal-poor GCs in the LE group (namely, NGC\,6254, NGC\,6544 and
NGC\,6809) cannot be ruled out.


In summary, from the point of view of kinematics, the low
energy defining this subgroup makes it hard to distinguish it from
the MD/MB subgroups (Fig.~\ref{kinematics}). On the basis
of the chemistry, while the metal-rich GCs NGC\,6441, Pal\,6 and
NGC\,6121 are clearly associated with the the MD/MB subgroups, the
association of the more metal-poor GCs NGC\,6254, NGC\,6544 and
NGC\,6809 is more uncertain, given that accreted and {\it in situ}
GCs occupy the same locus in the Si-Fe plane for those metallicities.
We conclude that, while the overal trend on the Si-Fe
plane of the LE GCs in our sample suggests an {\it in situ} origin,
the position of that subgroup in IOM space does not
preclude some of the members of that category having an accreted
origin.

%

Finally, we highlight the case of NGC~6388. Although that GC is
classified by \cite{Massari2019} as belonging to the MB subgroup, it
is characterized by very low [Si/Fe] ($\sim$--0.03), departing by
$\sim 2\sigma$ from the mean [Si/Fe] of that subgroup at
[Fe/H] $\sim$ --0.5. We discuss this GC in more detail in
Section~\ref{ngc6388}.

\subsection{Accreted subgroups} \label{accreted}

In this subsection, we examine the distribution of GCs of
accreted origin in the Si-Fe plane. 
Our analysis focuses on the following subgroups: H99,
GE, and Seq.

\begin{figure}
    \centering
    \includegraphics[width=1.\columnwidth]{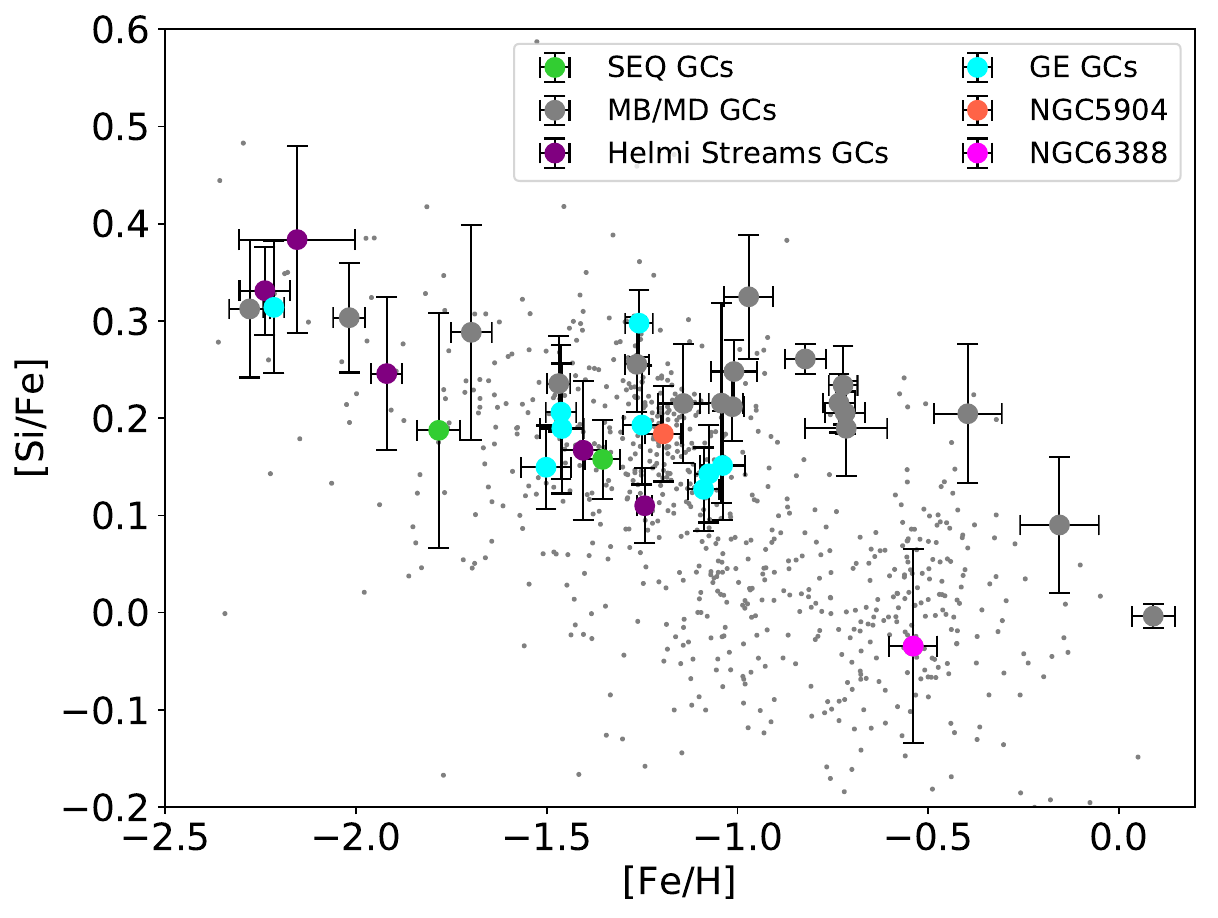}
    \caption{Mean [Si/Fe] vs [Fe/H] for the GE (cyan), Seq (green), H99 (purple) and MD/MB (grey) GC subgroups, illustrated alongside NGC\,5904 (red) and NGC\,6388 (magenta), with the 1 $\sigma$ spread represented in black error bars. In grey we show the halo field population defined as in \citet{Massari2019}. The GE, Seq and H99 accreted dwarf spheroidal subgroups occupy the same locus, displaying lower mean [Si/Fe] values to the GCs from the MD and MB populations at the same metallicity range --1.5 < [Fe/H] < --1. According to galaxy chemical-evolution models, this suggests that either: both accreted dwarf spheroidals must have had a similar chemical-evolution history and therefore have been of similar mass, or that some, possibly all, originate from the same accretion event.  Below [Fe/H] < --1.5, the \textit{in situ} and accreted groups are indistinguishable in the Si-Fe plane. NGC\,288 displays higher [Si/Fe] values than the rest of the GE subgroup GCs ($\sim$ 0.15 dex greater) of similar metallicity, however displays a clear accreted-like orbit (see Fig ~\ref{kinematics}). NGC\,5904 clearly occupies the same locus as the accreted population of GCs. However, due to the uncertainties in the measurements, it is impossible to suggest to which accreted subgroup NGC\,5904 belongs to. Along the same lines, NGC\,6388 occupies the same locus as the [Fe/H]-rich halo field population, which coupled with its retrograde orbit hints that this GC belongs to an accreted subgroup.}
    \label{SiFe_accreted}
\end{figure}

The data for these GCs are displayed in Fig ~\ref{SiFe_accreted}. We
first focus on a comparison between the positions occupied by the accreted
and {\it in situ} clusters.
Our results show that the GCs associated with the three putative
accreted systems all fall on the same locus in the [Si/Fe] plane,
positioned on average at lower [Si/Fe] values than the MD
and MB population (illustrated in Fig ~\ref{SiFe_accreted} as grey
points). This is commonly interpreted as the result of a history
of star formation and chemical enrichment typical of low-mass
galaxies, which differs from that of the Milky Way
\citep[e.g.,][]{Tolstoy2009}.  The accreted origin of the GCs that
are kinematically associated with GE, H99 and Seq is further confirmed
by the fact that their position in the [Si/Fe] plane mimics that
of field populations linked with past accretion events
(\citealp{Hayes2018,Mackereth2018}). We calculate the mean [Si/Fe] abundace given by GCs in the --1.5 < [Fe/H] < --1 regime for both our accreted and \textit{in situ} populations, and find that the accreted groups display on average [Si/Fe] = +0.17\pm 0.05, whereas the \textit{in situ} subgroups display a higher average abundance [Si/Fe] = +0.25\pm 0.03. This means that the two distributions differ at the $\sim$ 91.5\% level.


Having established that the accreted subgroups occupy a locus
of lower [Si/Fe] than that of {\it in situ} populations, we now
turn to a discussion of the relative positions of the GCs from the
three accreted subgroups in the Si-Fe space.  As pointed out above,
the GCs associated with the GE, H99, and Seq subgroups occupy the
same locus on the abundance plane, within the errors.  Such
similarity in chemical space can be understood in two possible ways.
On one hand, the different accreted subgroups may be associated to
three separate similar-mass satellites.  Alternatively, some,
or perhaps all of them, could be part of the same accreted satellite.
Consideration of the kinematic properties of the three systems may help
distinguish between these scenarios.  The GE system is strongly bound and
mildly retrograde, whereas the other two groups are slightly less bound,
with Seq being strongly retrograde and H99 strongly prograde.
\cite{Massari2019} argue that two of the GCs associated with the
Sequoia system (namely, NGC\,3201 and $\omega$~Cen) have a relatively
high probability of belonging to GE.  Moreover, they point out that the
Sequoia system's position in IOM space coincides with that
of debris that \cite{Helmi2018} ascribe to Gaia-Enceladus. 

Along the same lines, \citet{Massari2019} analysed the Helmi stream GCs employing the methodology described in \citet{Koppelman2019}, and found that, when accounting for the age uncertainties, H99 occupies a locus in age-metallicity space that is consistent with the \textit{Gaia}-Enceladus and Sequoia GC subgroups. Moreover, it has been shown that the field populations of the \textit{Gaia}-Enceladus, Sequoia, and Helmi Stream occupy the same locus in [Mg/Fe] vs [Fe/H] and [Al/Fe] vs [Fe/H] planes \citep{Koppelman2019b}, and that the Helmi Stream displays an MDF which peaks at a value of [Fe/H] $\sim$ --1.5 \citep{Koppelman2019}, similar to value at which the Gaia-Enceladus MDF peaks (\citealp{Helmi2018,Mackereth2018}).

We
conclude that the combined evidence from GC subgroup chemistry and
kinematics is suggestive of either a common origin for the Sequoia, Helmi Stream and
Gaia-Enceladus stellar systems, or that these subgroups are associated with satellites which underwent similar chemical enrichment histories.\\



In closing this subsection we comment on the interesting case
of NGC~288.  On the basis of kinematics, \cite{Massari2019} assign
it unambiguously to the GE subgroup, with a retrograde orbit and
high orbital energy (see Fig~\ref{kinematics}).  However, its
elemental abundances place it squarely on the {\it in situ} branch,
$\sim 2 \sigma$ off the mean of the GE subgroup at the same [Fe/H]. We checked to see if this result survives when other $\alpha$-elements are considered, and find that NGC\,288 also displays [Mg/Fe] and [Ca/Fe] values $\sim$ 2 $\sigma$ away from the mean of the GE subgroup, with the GE subgroup presenting mean values of  $\langle$[Mg/Fe]$\rangle_{GE}$ = +0.17 \pm 0.07 and  $\langle$[Ca/Fe]$\rangle_{GE}$ = +0.19 \pm 0.04, respectively, and NGC\,288 displaying [Mg/Fe]$_{NGC\,288}$ = +0.31 \pm 0.04 and [Ca/Fe]$_{NGC\,288}$ = +0.26 \pm 0.09, for the same [Fe/H]. It is difficult to reconcile the orbital and chemical properties of NGC\,288, so we suggest that NGC\,288 is likely an accreted GC with a peculiar chemical composition. 

\subsection{NGC 5904 and Liller 1}
\label{unclassified}

The study by \cite{Massari2019} did not assign Liller\,1 to any
particular kinematic subgroup, and concluded that NGC\,5904 could be associated with either the GE or H99 subgroup. In this subsection we examine these
GCs' positions in chemical space to see whether that information can help clarify whether they have an accreted or
\textit{in situ} origin.

We first compare the [Si/Fe] vs [Fe/H] abundance measurements
obtained for Liller\,1 with the other identified subgroups (see
Fig.~\ref{SiFe_insitu}).  Our results show that within the uncertainties,
Liller\,1 occupies the same locus in the [Si/Fe] vs [Fe/H] abundance
plane as the \textit{in situ} population, and therefore belongs to either the MD or MB subgroup. 
Unfortunately, there is no 6D phase-space information for Liller 1 \citep[see][for details]{Vasiliev2019}, and therefore we are unable to place kinematic constrains on the origin of this GC. Furthermore, Liller\,1 is quite
metal rich, with a mean value of <[Fe/H]>$_{Liller\,1}\simeq$
--0.03\pm --0.05, which is much higher than the metallicities of the accreted GCs. Thus, our results
suggest that  Liller\,1 has an \textit{in situ} origin, agreeing with previous
studies \citep[e.g.][]{Bica2016}. \\

In the case of NGC~5904, our results show that the mean abundances place it on the same locus as the GCs associated
to the accreted subgroups.  Therefore, within
the uncertainties our results suggest that NGC\,5904 has an accreted
origin \citep[agreeing with the suggestion by][]{Massari2019}. However, since it is impossible to distinguish the accreted groups in Si-Fe space, we cannot establish an association of NGC~5904 to any particular accreted subgroup.

\subsection{NGC 6388}
\label{ngc6388}


As pointed out in Section~\ref{DiscBulge}, NGC\,6388 displays a very low [Si/Fe] abundance ratio, departing significantly from the locus of
the MB subgroup, to which it was associated by \cite{Massari2019}.
Its position on the Si-Fe plane is consistent with an
extrapolation towards high metallicity of the trend established by
the accreted subgroups at [Fe/H] $\simless$ --1.  It also falls on
top of the accreted field population in Fig.~\ref{SiFe_accreted}.
The mean abundances for NGC~6388 are based on values for 24 members,
which we consider to be statistically robust. Specifically, the mean silicon abundance of NGC\,6388 members (<[Si/Fe]> = --0.03\pm0.1) deviates from that of the high $\alpha$ sequence at same [Fe/H] (<[Si/Fe]>$_{high \alpha}$ = +0.17\pm0.05) by $\sim 2~\sigma$.  It is also lower than that of the low $\alpha$ sequence  (<[Si/Fe]>$_{low \alpha}$ = +0.02\pm0.04) by $\sim 1~\sigma$. We note, however,
that \citet{Carreta2018} obtained abundances for a comparable
sample of NGC~6388 members, obtaining $\sim$0.4 dex higher mean [Si/Fe].
\cite{Wallerstein2007} also obtained a $\sim$0.3 dex higher mean
[Si/Fe], although their mean abundances of Ti and Ca were around
solar or lower ($\sim$+0.06 and --0.05, respectively), depending
on the $\log g$ adopted. On the other hand, \citet{Meszaros2019} analysed the APOGEE spectra using a different pipeline,
obtaining similar results to those presented in this paper.

In order to check whether our result is due to systematics
in the [Si/Fe] abundance ratios of NGC~6388 stars, we examined the
abundances of other $\alpha$-elements, such as Mg and Ca.  For the
latter element we found $\langle$[Ca/Fe]$\rangle$ = +0.09 \pm 0.11, which is lower than the
values for the MB population at the same [Fe/H] ([Ca/Fe] = +0.19 \pm 0.02), deviating at the $\sim$ 1~$\sigma$ level. Before estimating mean [Mg/Fe], one
needs to select GC members that are not affected by the multiple population
phenomenon.  In order to isolate NGC\,6388 stars belonging to the so-called
``first population'', we proceeded as follows. We use [N/Fe] in order to identify first population stars, since this abundance ratio is strongly enhanced in their second population counterparts (\citealp[see, e.g.][]{Renzini2015,Schiavon2017_nrich,Schiavon2017,Bastian2018}). We define as first population stars those  located within the bottom quartile of the [N/Fe] distribution of NGC~6388 members.  
By proceeding in this way, we are confident that we managed to isolate a subsample of first population GC stars, whose Mg abundances are not affected by the multiple populations phenomenon.
For this subsample, we obtained
$\langle$[Mg/Fe]$\rangle$ = +0.07 \pm 0.11, which again is lower than the mean value for
the MB population ($\langle$[Mg/Fe]$\rangle$ = +0.27 \pm 0.06) by $\sim$ 2 $\sigma$. 

It is worth noticing that the relative position of NGC 6388 in $\alpha$-Fe space is not the same according to different $\alpha$-elements.  When Si is considered, NGC\,6388 falls below the low-$\alpha$ sequence at the 1 $\sigma$ level.  On the other hand, the cluster falls on top of the low-$\alpha$ sequence when Mg or Ca are considered.

Due to NGC\,6388 being a bulge GC, positioned in a crowded and dense field, it is likely that our sample is contaminated by field stars, mainly in the GC foreground. To ensure our previous findings are robust, we minimise field contamination by considering only N-rich stars, which belong to the "second-population" GC population stars. To obtain a clean sample of "second population" stars, we select only stars located at the top quartile of the [N/Fe] distribution. For second-population NGC\,6388 stars defined in that manner, we find an average of $\langle$ [Si/Fe] $\rangle$ = --0.07 \pm 0.08, which places NGC\,6388 even further away from the {\it in situ} population. This solidifies our initial findings, and confirms that NGC\,6388 displays lower [Si/Fe] abundances than those of other MB GCs of similar metallicity. 

\cite{Myeong2019} studied the properties of NGC~6388 in detail,
showing that, on one hand, it is consistent with an accreted
origin on account of its kinematic properties, but on the other its
combination of age and metallicity places it on top of the relation
defined by the {\it in situ} GC population for those two parameters
\citep[see also][]{Kruijssen2019a}.  We determined the orbital energy and
azimuthal action of NGC~6388 (see Fig ~\ref{kinematics}), finding its orbit
to be retrograde, in agreement with \citet{Myeong2019}, but cannot
distinguish between a possible association to the MB, LE, or the Seq
subgroups. Furthermore, \citet[][]{Milone2019} classified NGC\,6388 as a Type II GC, based on the ratio of first-population to second-population stars. They also obtain the IOM of this GC, and conclude that NGC\,6388 is likely from an accreted origin.

We summarise our results for NGC\,6388 as follows: 1) the [Si/Fe] abundance for this GC differs from that of the MB/MD population at the 2~$\sigma$ level; 2) it differs from that of the low-$\alpha$ at the 1~$\sigma$ level; 3) Figure \ref{SiFe_accreted} shows that NGC\,6388 falls on top of the accreted {\it field} halo populations in the Si-Fe plane; 4) its position in the IOM does not provide a unique distinction between an accreted or {\it in situ} origin.  In view of these results, it is fair to conclude that the data suggest a possible accreted origin for NGC\,6388.

\section{Conclusions}
\label{conclusion}

In this work, we have employed the sixteenth data release from the
SDSS/APOGEE survey in order to map the kinematic properties of
Galactic GCs into their positions in the chemical compositions
space.  We contrast positions, and APOGEE abundances and radial
velocities with information gathered from the 2010 edition
of the Harris GC catalogue \citep{Harris1996} and the Baumgardt
$\&$ Hilker GC catalogue \citep{Baumgardt2018,Baumgardt2019} to
obtain a primary GC sample in APOGEE, which we refine to obtain an
accurate GC membership list. We obtain a final main GC sample of
3090 stars, associated with 46 GCs, from which then $\omega$~Cen, Terzan\,5 and NGC\,6715 are removed for reasons detailed in Section \ref{main_sampling}, leaving us with a sample of 1,728 stars associated with 43 GCs. We assign membership to various kinematic subgroups
according to the classification by
\citet{Massari2019}. We then examine
the distributions of the various GC subgroups in chemical space,
more specifically the plane defined by $\alpha$ and Fe abundances.
After excluding GCs with fewer than three member stars, we identify
in our sample 9 GCs associated to the MD group, 10 to the MB, 9 to
the GE dwarf spheroidal, 5 to the H99, 6 to the LE group,
2 to the Seq dwarf spheroidal and 0 to the Sag dwarf spheroidal.
Furthermore, we find 2 GCs (namely, Liller\,1 and NGC\,5904) for
which there remains an uncertain association.

We make use of Si abundance measurements in APOGEE as our tracer
of $\alpha$-elements abundance, and plot them as a function of
[Fe/H] with the goal of gaining insight into the nature of the
different kinematic subgroups. In this comparison, we search for
any possible plateau or knees that may present themselves in an
[$\alpha$/Fe] vs [Fe/H] plane. Our results and conclusions are unchanged
by adoption of the sample and abundances presented by \citet{Meszaros2019}.
Our conclusions can be summarised as follows:

\begin{enumerate}

\item  When considered together the {\it in situ} GC subgroups
(Main Disk and Main Bulge, MD and MB) and the low-energy group (LE)
follow the overall trend of the {\it in situ} populations (MB and
MD) in chemical space, with a [Si/Fe]$\sim$+0.25 plateau at low
metallicity and a change of slope (so-called ``knee'') at
[Fe/H]$\sim$--0.8.

\item GCs from accreted subgroups, namely Gaia-Enceladus (GE),
Helmi streams (H99), and Sequoia (Seq) fall on the same area of
chemical space as accreted field populations.  This locus is
characterized by [Si/Fe]$\simless$ + 0.2 at
--1.5 $\simless$[Fe/H]$\simless$ --1.0, going down to solar or near
sub-solar [Si/Fe] for [Fe/H]$\sim$ --0.5.  At [Fe/H]$\simless$ --1.5, GCs
from the accreted and {\it in situ} subgroups are indistinguishable in the Si-Fe
plane.

\item When examined separately, the MD, MB, and LE subgroups track
the field population, however due to the relatively small sample size
these subgroups do not sample the metallicity space densely enough
to define the trend separately from the other subgroups.  Three out
of six of the LE GCs (namely, NGC\,6121, NGC\,6441 and Pal\,6) fall on the
high-metallicity side of the knee and follow the trend of the field
populations, leading to the conclusion that they have an {\it in situ} origin.  The three metal-poor GCs from the LE subgroup (namely, NGC\,6254, NGC\,6544 and NGC\,6809)
fall in the region of Si-Fe where accreted and {\it in situ} GCs
are indistinguishable, so their origin is less certain.  We conclude that
the chemical properties of the LE subgroup as a whole are consistent with
an {\it in situ} origin, but given its borderline position in IOM space, individual clusters belonging to this subgroup could have an
accreted origin.

\item GCs from the accreted H99 and GE subgroups occupy the same
position in chemical space.  That is also the case for the GCs in the Seq
group, but since our sample contains only two Seq GCs, the result for that
subgroup is not as firm.  This result suggests that GCs from these
subgroups are associated to accreted satellites of similar masses, or possibly originating from one common progenitor.  Based
on its position on the IOM space, it is possible that the
GCs from the Seq and GE subgroups actually once belonged to the same
system, as suggested by other groups \citep[e.g.,][]{Massari2019}.


\item NGC~6388 is found to present Si, Mg, and Ca abundances
that are considerably lower than those of other GCs in the main
bulge subgroup and similar [Fe/H]. The evidence from other
studies in the literature is not conclusive, so more studies exploring
other spectral regions and different $\alpha$-elements are required
to ascertain the low-$\alpha$ nature of this GC.  Considering the orbital characteristics, a confirmation of this result
will lend strong support to the notion that NGC~6388 was in fact
accreted to the Milky Way, as also suggested by other groups (\citealp[e.g.][]{Milone2019,Myeong2019}).


\item NGC\,288 is found to present Si, Ca and Mg abundances that
are considerably higher than those of other accreted GCs of
similar [Fe/H]. It is characterized by
a highly unbound retrograde orbit. We conclude that NGC\,288 is an unusual GC where the kinematic
properties suggest an accreted origin which is not fully compatible with
the its chemistry.  More work is needed to clarify the origin of this
object.


\item Comparison of the mean [Si/Fe] vs [Fe/H] chemical compositions of
Liller\,1 and NGC\,5904 with those of the different kinematic
subgroups suggests that Liller\,1  possibly associated with the \textit{in situ} subgroups and that NGC\,5904
was likely accreted.  We cannot, on the basis of the extant data, establish to which accreted subgroup this GC is associated.

\end{enumerate}

In summary, the information provided by the sixteenth APOGEE data
release has enabled a study of the chemical-abundances of the
Galactic GC system, shedding light on the origin of a reasonably
representative sample of the Milky Way GC system. The combination
of the chemical-abundance information delivered by APOGEE and the
kinematic 6D phase-space information provided by Gaia provides
interesting insights into the origin of the Milky Way GCs.
Expansion of such data bases to a larger sample of Galactic GCs
will shed new light on the Galaxy's early mass assembly history.

\section*{Acknowledgements}

The authors thank Vasily Belokurov and Dan Perley for helpful discussions, and
Nate Bastian, Joel Pfeffer, Phil James and Diederik Kruijssen for
comments on an early version of the manuscript.  DHD thanks Sue,
Alex and Debra for their moral support.  He is funded by the Science
and Technologies Facilities Council (STFC, UK) studentship. JTM
acknowledges support from the ERC Consolidator Grant funding scheme
(project ASTEROCHRONOMETRY, G.A.  n. 772293). H.J. acknowledges
support from the Crafoord Foundation, Stiftelsen Olle Engkvist
Byggm\"astare, and Ruth och Nils-Erik Stenb\"acks stiftelse. P.M.F.
acknowledges support for this research from the National Science
Foundation (AST-1715662). T.C.B. acknowledges partial support from grant PHY 14-30152; Physics Frontier Center / JINA Center for the Evolution of the Elements (JINA-CEE),
awarded by the U.S. National Science Foundation. D.G gratefully ackowledges support from the Chilean Centro de Exelencia en Astro\'{i}sica y Tecnolog\'{i}as Afines (CATA) BASAL grant AFB-170002. D.G. also ackowledges financial support from the Direcci\'{o}n de Investigaci\'{o}n y Desarrollo de la Universidad de la Serena through the Programa de Incentivo a la Investigaci'{o}n de Acad\'{e}micos (PIA-DIDULS). SzM has been supported by the Premium Postdoctoral
Research Program and J{\'a}nos Bolyai Research Scholarship of the Hungarian Academy of
Sciences, by the Hungarian NKFI Grants K-119517 and GINOP-2.3.2-15-2016-00003 of the Hungarian National
Research, Development and Innovation Office. J.G.F-T is supported by FONDECYT No. 3180210 and Becas Iberoam\'erica Investigador 2019, Banco Santander Chile.

Funding for the Sloan Digital Sky Survey IV has been provided by
the Alfred P. Sloan Foundation, the U.S. Department of Energy Office
of Science, and the Participating Institutions. SDSS acknowledges
support and resources from the Center for High-Performance Computing
at the University of Utah. The SDSS web site is www.sdss.org. SDSS
is managed by the Astrophysical Research Consortium for the
Participating Institutions of the SDSS Collaboration including the
Brazilian Participation Group, the Carnegie Institution for Science,
Carnegie Mellon University, the Chilean Participation Group, the
French Participation Group, Harvard-Smithsonian Center for Astrophysics,
Instituto de Astrof\'{i}sica de Canarias, The Johns Hopkins University,
Kavli Institute for the Physics and Mathematics of the Universe
(IPMU) / University of Tokyo, the Korean Participation Group,
Lawrence Berkeley National Laboratory, Leibniz Institut f\"{u}r Astrophysik
Potsdam (AIP), Max-Planck-Institut f\"{u}r Astronomie (MPIA Heidelberg),
Max-Planck-Institut f\"{u}r Astrophysik (MPA Garching), Max-Planck-Institut
f\"{u}r Extraterrestrische Physik (MPE), National Astronomical Observatories
of China, New Mexico State University, New York University, University
of Notre Dame, Observatório Nacional / MCTI, The Ohio State University,
Pennsylvania State University, Shanghai Astronomical Observatory,
United Kingdom Participation Group, Universidad Nacional Autónoma
de México, University of Arizona, University of Colorado Boulder,
University of Oxford, University of Portsmouth, University of Utah,
University of Virginia, University of Washington, University of
Wisconsin, Vanderbilt University, and Yale University.




\bibliographystyle{mnras}
\bibliography{ref}

\appendix
\section{GC metallicity and radial velocitydistribution functions}
\clearpage
\label{MDFS}
\begin{figure*}
\center
    \includegraphics[width=1\textwidth,height=0.87\textheight]{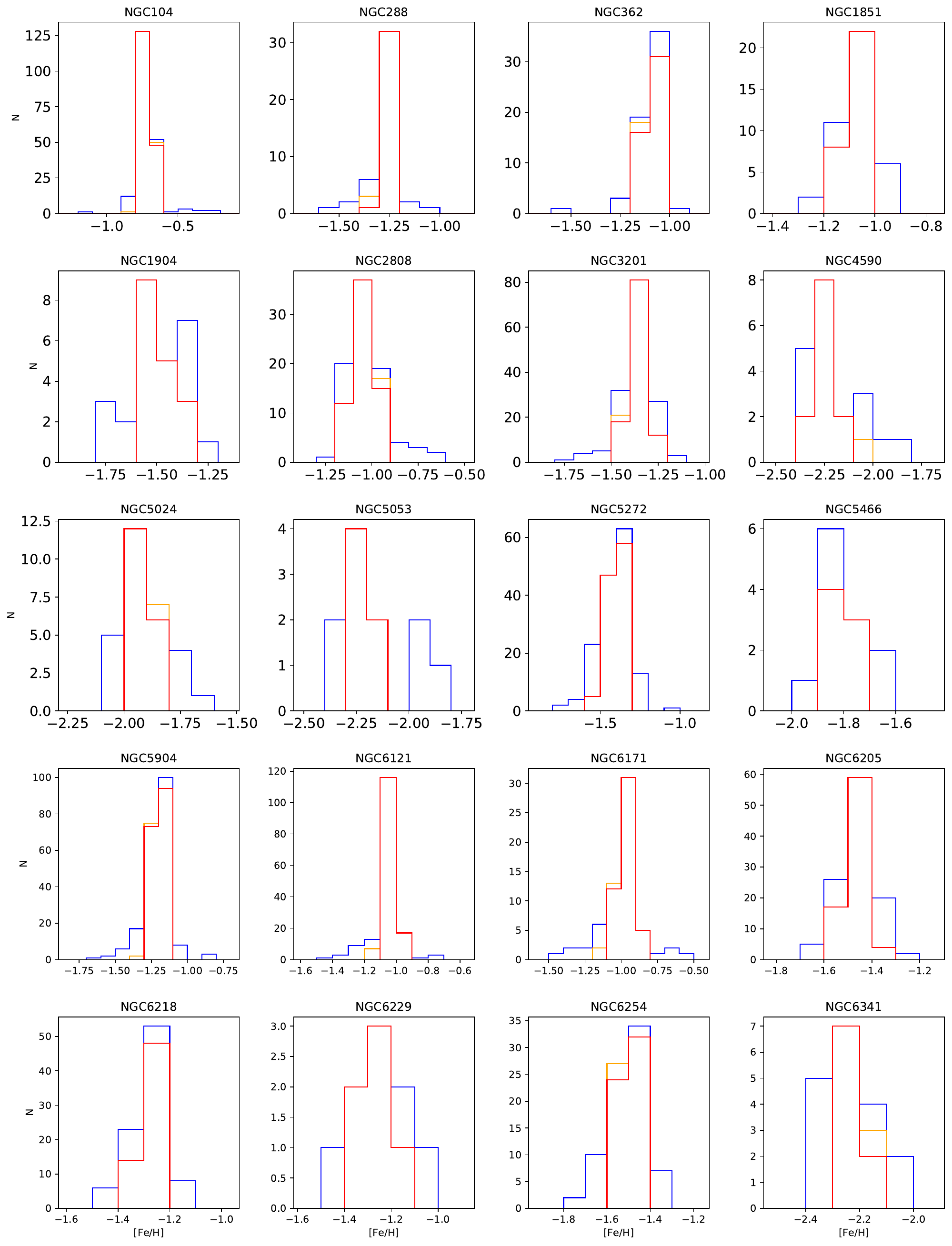}
    \caption{Metallicity distribution functions for the GCs in the main sample. The blue histogram represents the GC members obtained before employing the MDF sigma clip cut, for which the resulting members are highlighted as the yellow histogram. The red histogram are the resulting members after performing a second MDF sigma clip. Recall that for each GC a different clip was applied, depending on the cluster and the MDF distribution.}.
    \label{mdfs}
\end{figure*}
\clearpage
\begin{figure*}
    \includegraphics[width=1\textwidth,height=0.87\textheight]{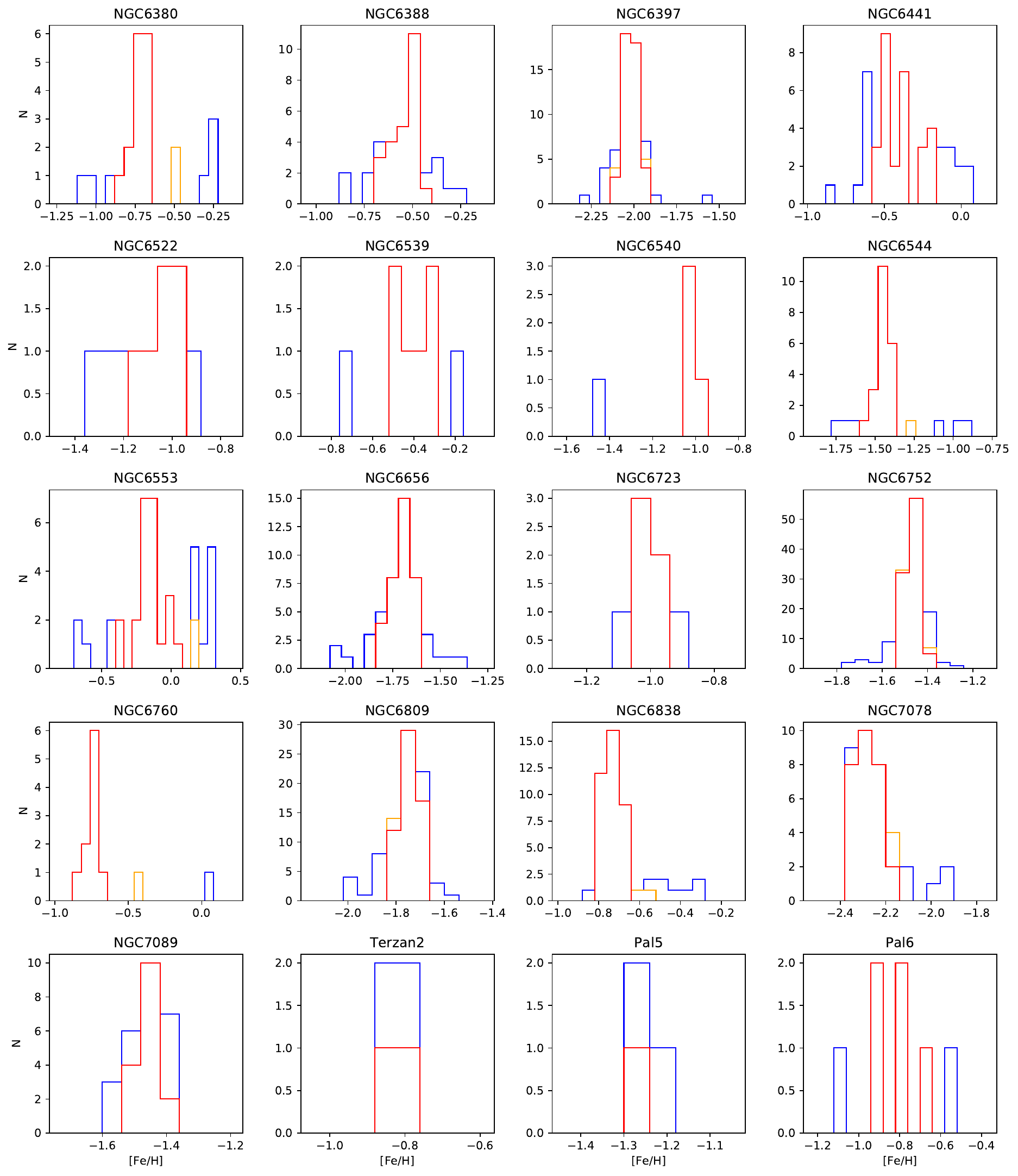}
    \caption{Fig ~\ref{mdfs} continued.}
\end{figure*}
\clearpage
\begin{figure}
    \includegraphics[width=0.75\textwidth,height=0.2\textheight]{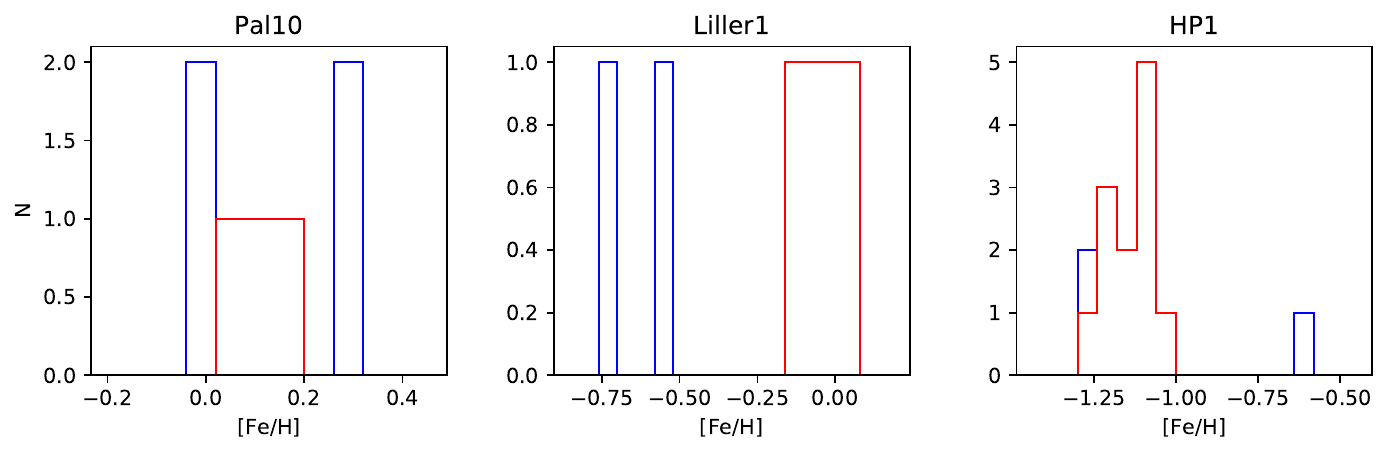}
\caption{Fig ~\ref{mdfs} continued.}
\end{figure}
\begin{figure*}
\center
    \includegraphics[width=1\textwidth,height=0.87\textheight]{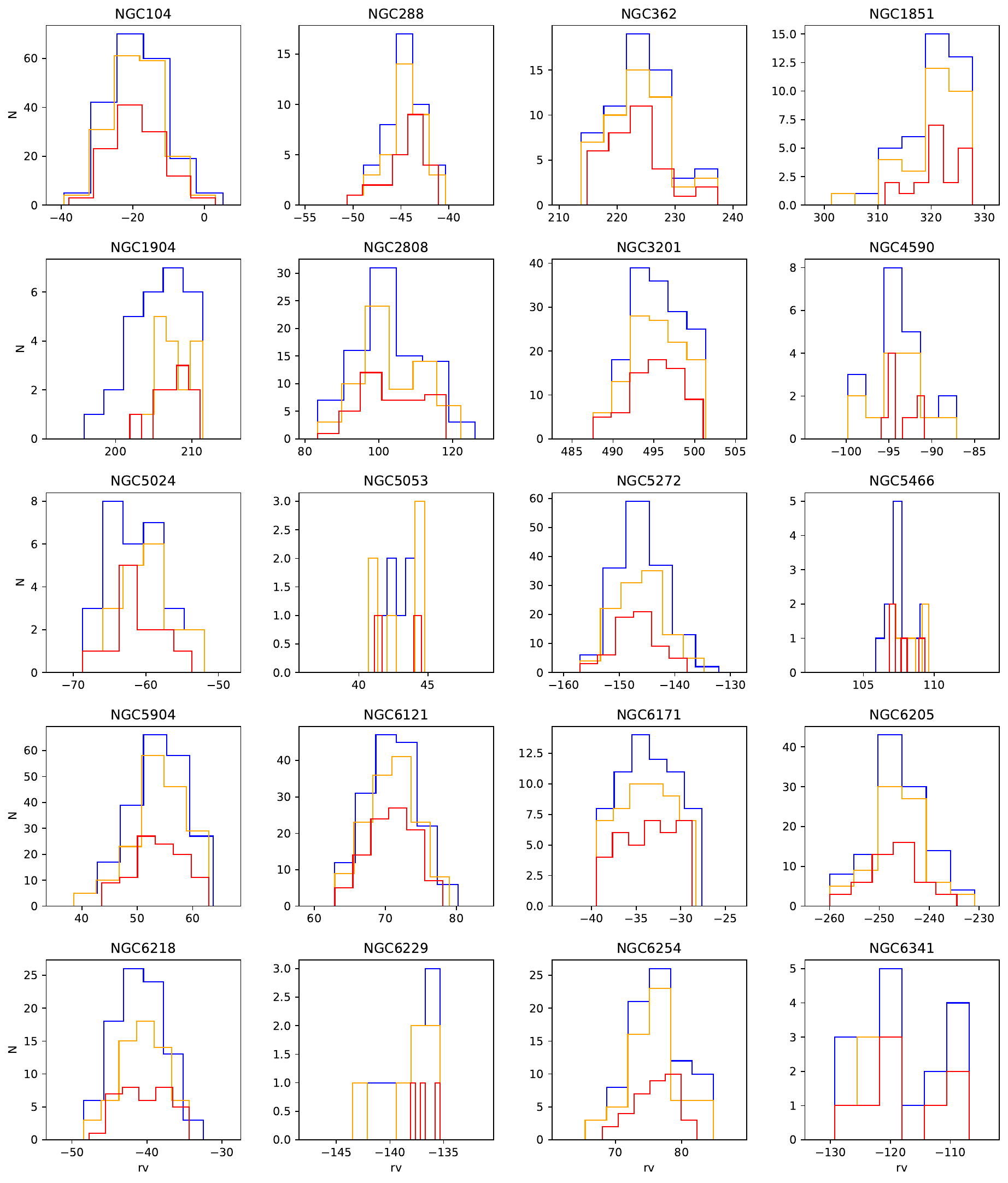}
    \caption{Radial velocities for the GCs in the main sample. The blue histogram represents the GC members obtained before employing the MDF sigma clip cut, for which the resulting members are highlighted as the yellow histogram. The red histogram are the resulting members after performing a second MDF sigma clip.}
    \label{radvels}
\end{figure*}
\clearpage
\begin{figure}
    \includegraphics[width=1\textwidth,height=0.87\textheight]{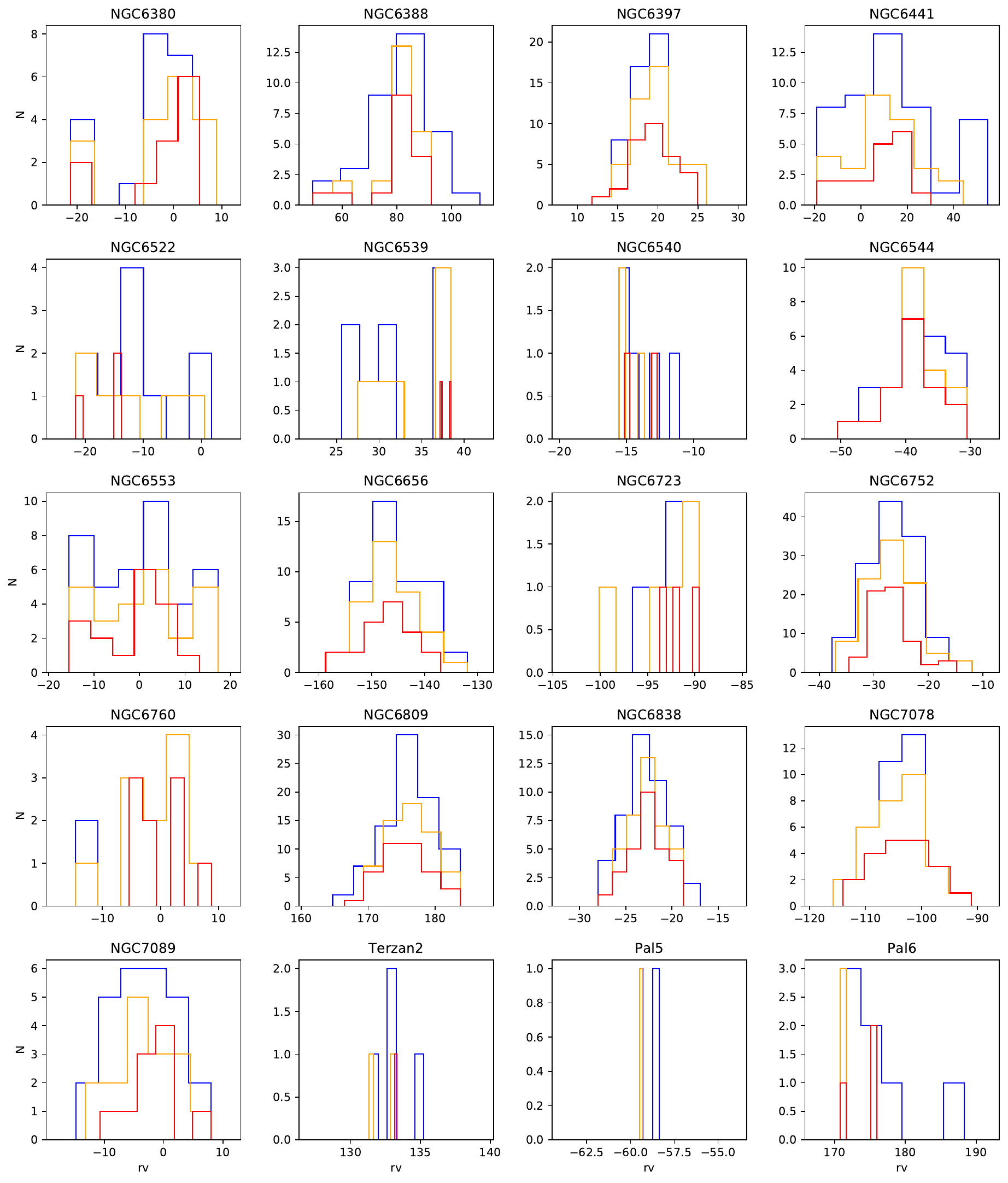}
    \caption{Fig ~\ref{radvels} continued.}
\end{figure}
\clearpage
\begin{figure}
    \includegraphics[width=0.75\textwidth,height=0.2\textheight]{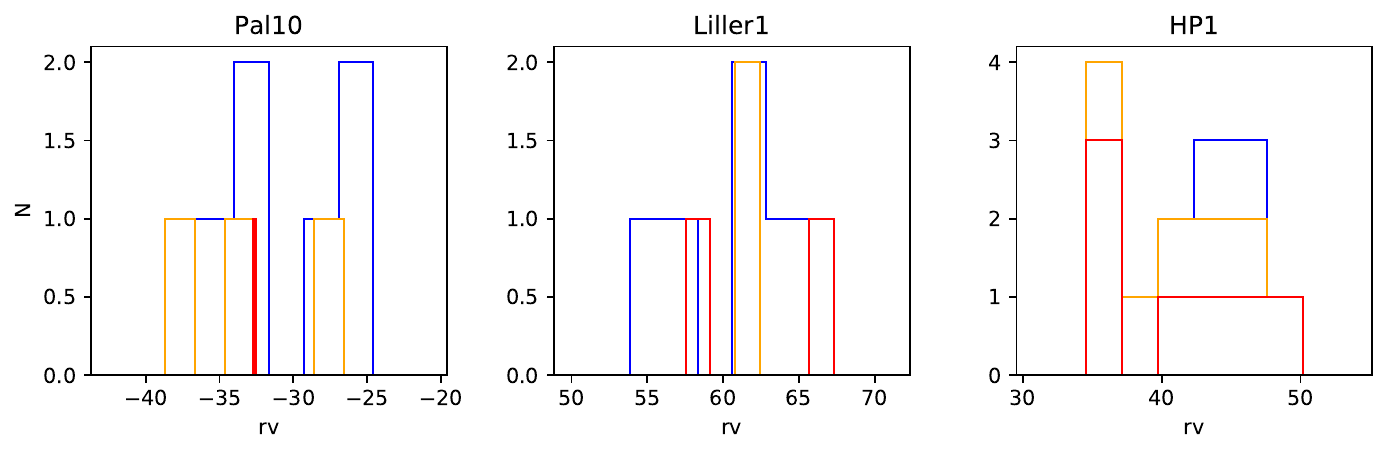}
\caption{Fig ~\ref{radvels} continued.}
\end{figure}

\bsp	
\label{lastpage}
\end{document}